\documentclass{llncs}
\usepackage{amsmath,amsgen,amstext,amssymb}
\usepackage{amsfonts}
\usepackage{latexsym}
\usepackage{graphicx}
\usepackage{epsfig}
\usepackage{subfigure}
\usepackage{longtable}
\usepackage{xspace}

\usepackage{booktabs}

\graphicspath{{img/}}

\newcommand{\beq}{\begin{equation}}
\newcommand{\deq}{\end{equation}}

\newcommand{\beqm}{\begin{equation*}}
\newcommand{\deqm}{\end{equation*}}

\newcommand{\baq}{\begin{eqnarray}}
\newcommand{\daq}{\end{eqnarray}}

\newcommand{\baqm}{\begin{eqnarray*}}
\newcommand{\daqm}{\end{eqnarray*}}

\newcommand{\mqed}{\hfill $\square$}

\newcommand{\eps}{\varepsilon}

\newcommand{\co}{\ensuremath{\mathcal{O}}}
\newcommand{\acfbet}{$\alpha$-CF betweenness\xspace}

\newtheorem{thm}{Theorem}

\title{Alpha current flow betweenness centrality\thanks{This research is partially funded by
Inria Alcatel-Lucent Joint Lab, by the European Commission within
the framework of the CONGAS project FP7-ICT-2011-8-317672, see www.congas-project.eu, and by the EU-FET Open grant NADINE (288956).}}

\author{Konstantin Avrachenkov$^1$, Nelly Litvak$^2$, Vasily Medyanikov$^3$, Marina Sokol$^1$}

\institute{Inria Sophia Antipolis, 2004 Route des Lucioles, Sophia-Antipolis, France\and University of Twente, P.O.Box 217, 7500AE, Enschede, The Netherlands\and
St. Petersburg State University, 7-9, Universitetskaya nab., St. Petersburg, Russia}

\begin{document}

\maketitle

\begin{abstract}
A class of centrality measures called betweenness centralities reflects degree of participation of edges or nodes in communication between different parts of the network. The original shortest-path betweenness centrality is based
on counting shortest paths which go through a node or an edge. One of shortcomings of the shortest-path betweenness centrality is that it ignores the paths that might be one or two steps longer than the shortest paths, while the edges
on such paths can be important for communication processes in the network. To rectify this shortcoming a current flow
betweenness centrality has been proposed. Similarly to the shortest path betwe
has prohibitive complexity for large size networks. In the present work we propose two regularizations of the current
flow betweenness centrality, $\alpha$-current flow betweenness and truncated $\alpha$-current flow betweenness,
which can be computed fast and correlate well with the original current flow betweenness.
\end{abstract}

\section{Introduction}
	A class of centrality measures called betweenness centralities reflects degree of participation of edges or nodes in communication between different parts of the network.
%	\footnote{Here we will introduce various betweenness centralities for network edges. Analogous notions for the nodes are very similar and details can be found in corresponding references.}
	The first notion of betweenness centrality was introduced by Freeman \cite{Freeman1977}. Let $s, t \in V$ be a pair of nodes in an undirected network $G = (V, E)$. We denote $|V|=n$, $|E|=m$, and let $d_v$ be the degree of node $v$. Let $\sigma_{s, t}$ be the number of shortest paths connecting nodes $s$ and $t$ and denote $\sigma_{s, t}(e)$ the number of shortest path connecting nodes $s$ and $t$ passing through edge $e$. Then betweenness centrality of edge $e$ is calculated as follows:
	\begin{equation}
	\label{eq:sp-betweenness}C_\text{B}(e) = \frac{1}{n(n - 1)} \sum_{s, t \in V} { \frac{\sigma_{s, t}(e)} {\sigma_{s, t}} }\end{equation}
Computational complexity of the best known algorithm for computing the betweenness in (\ref{eq:sp-betweenness})is $\co(mn)$ \cite{Brandes2001}. This limits its applicability for large graphs.
	
%In order to reduce the computational costs, the $k$-betweenness was proposed as an approximation of the original expression (\ref{eq:sp-betweenness}), see \cite{Ercsey-Ravasz2010,Gkorou,Pfeffer2012}. This approximation includes only paths of length at most $k$:
%	$$C_\text{k-B}(e) = \frac{1}{n(n - 1)} \sum_{s, t: d(s, t) <= k} { \frac{\sigma_{s, t}(v)} {\sigma_{s, t}} }$$
%	The advantage of this measure is that it can be computed fast if $k$ considerably smaller than the average distance. By varying the parameter $k$ we can obtain the whole spectrum of centrality measures from very local (taking into account just degrees of the nodes adjacent to the edge) to the global original betweenness centrality.

One of shortcomings of the betweenness centrality in (\ref{eq:sp-betweenness})is that it takes into accounts only the shortest paths, ignoring the paths that might be one or two steps longer, while the edges on such paths can be important for communication processes in the network. In order to take such paths into account, Newman~\cite{Newmana} and Brandes and Fleischer~\cite{Brandes2005a} introduced the current flow betweenness centrality (CF-betweenness). In \cite{Newmana,Brandes2005a} the graph is regarded as an electrical network with edges being unit resistances. The CF-betweenness of an edge is the amount of current that flows through it, averaged over all source-destination pairs, when one unit of current is induced at the source, and the destination (sink) is connected to the ground. This exploits the well known relation between electrical networks and reversible Markov chains, see e.g.
\cite{Aldous-Fill,DoyleSnell}.

The computational difficulty of Betweenness and the CF-betweenness is that the computations must be done over the set of all source-destination pairs. The best previously known computational complexity for the CF-betweenness is $\co(I(n-1) + mn\log n)$ where $I(n-1)$ is the complexity of the inversion of matrix of dimension $n-1$.
	
	In the present work we introduce new betweenness centrality measures: $\alpha$-current flow betweenness (\acfbet) and its truncated version. The main purpose of these new measures is to bring down the high cost of the CF-flow betweenness computation. Our proposed measures are very close in performance to the CF-betweenness, but they are comparable to the PageRank algorithm~\cite{Brin1998107} in their modest computational complexity. 	
	%	The presented list of betweenness measures is not exhaustive ant there exist other approximations and generalizations, see \cite{Kang}, \cite{Alahakoon}, \cite{Jiang2009}, \cite{Chan2009}.
	Our goal is to provide and analyze efficient algorithms for \acfbet
and truncated \acfbet, to compare the \acfbet to other centrality measures.
	 	
%In Section~\ref{sec:acfbet} we introduce the definition of \acfbet, present the algorithm for its computation and derive its computational complexity. In Section~\ref{sec:truncated} we present a truncated version of the \acfbet that is better correlated with the CF-betweennes when $\alpha$ is not very close to one. In Section~\ref{sec:test} we describe our datasets. We further study the properties of \acfbet on three smaller datasets, on which the original CF-betweennes can be computed. In Section~\ref{sec:correlations} we consider the correlations between \acfbet and other measures. In Section~\ref{sec:robust} we demonstrate that our new measure adequately finds the nodes that are essential for the network connectivity. The larger graph, for which the original CF-betweenness cannot be computed, is analyzed in Section~\ref{sec:enron}. We conclude with discussion in Section~\ref{sec:discussion}.

\section{Alpha current flow betweenness}
\label{sec:acfbet}
We view the graph $G$ as an electrical network where each edge has resistance $1/\alpha$, and each node is connected to
ground node $n+1$ by an edge with resistance $1/(1-\alpha)$. This is in the spirit of the PageRank, indeed, the current (probability flow) is inversely proportional to the resistance,
and thus the fraction $\alpha$ of the current from a node flows to the network, while the fraction $(1-\alpha)$ of the current is directed to the sink. Since the graph is undirected, we use a convention that $(v,w)$ and $(w,v)$ represent the same arc in $E$, but depending on the chosen direction the current along this arc is considered to be positive or negative.

Assume that a unit of current is supplied to a source node $s\in V$, and there is a destination node $t\in V$ connected to the ground. Let $\varphi_v^{(s,t)}$ denote the absolute potential of node $v\in V$, if $s$ is a source $s$, and $t$ is the destination. Assume without loss
of generality that $s=1$ and $t=n$ ($\varphi^{(1,n)}_n=\varphi^{(1,n)}_{n+1}=0$).
The vector of absolute potentials of the other nodes $\varphi^{(1,n)}=[\varphi^{(1,n)}_1,...,\varphi^{(1,n)}_{n-1}]^T$
is a solution of the following system of equations (Kirchhoff's current law):
\beq
\label{eq:KCL}
[\tilde D - \alpha \tilde A] \varphi^{(1,n)} = \tilde b,
\deq
where $\tilde D$ and $\tilde A$ are the degree and adjacency matrices of the graph without node $n$
and $\tilde b = [1,0,...,0]^T$, see \cite{Brandes2005a}.

Obviously, we would not like to solve a separate linear system for each source-destination pair with
different left hand side coefficient matrix $[\tilde D - \alpha \tilde A]$. In the following
theorem we demonstrate that we need to only invert the coefficient matrix $[D-\alpha A]$.

\begin{thm}\label{thm:Cmatrix}
The voltage drop along the edge $(v,w)$ is given by
\begin{equation}\label{eq:potdifmodif}
\varphi^{(s,t)}_v-\varphi^{(s,t)}_w = (c_{s,v}-c_{s,w}) + \frac{c_{s,t}}{c_{t,t}} (c_{t,w}-c_{t,v}),
\end{equation}
where $(c_{v,w})_{v,w\in V}$, are the elements of the matrix $C=[D - \alpha A]^{-1}$.
\end{thm}

\noindent {\bf Proof:} Assume again without loss
of generality that $s=1$ and $t=n$. The matrix $[D-\alpha A]$ can be written in the following block structure
$$
D - \alpha A =
\left[\begin{array}{cc}
\tilde D - \alpha \tilde A & -\alpha \tilde a\\
-\alpha \tilde a^T & d_n
\end{array}\right],
\quad \mbox{with} \quad
\tilde a =
\left[\begin{array}{c}
a_{1,n}\\
a_{2,n}\\
\vdots\\
a_{n-1,n}
\end{array}\right].
$$
Then, divide accordingly the elements of the inverse matrix
$$
C=[D - \alpha A]^{-1}
=\left[\begin{array}{cc}
\tilde C & \tilde c\\
\tilde c^T & c_{n,n}
\end{array}\right].
$$
Writing the relation $[D - \alpha A]C=I$ in the block form yields
\beq
\label{eq:InvBlock1}
[\tilde D - \alpha \tilde A] \tilde C - \alpha \tilde a \tilde c^T = I,
\deq
\beq
\label{eq:InvBlock2}
[\tilde D - \alpha \tilde A] \tilde c - \alpha \tilde a \tilde c_{n,n} = 0.
\deq
Premultiplying equation (\ref{eq:InvBlock1}) by $[\tilde D - \alpha \tilde A]^{-1}$,
we obtain
\beq
\label{eq:InvBlock1a}
[\tilde D - \alpha \tilde A]^{-1} = \tilde C - \alpha [\tilde D - \alpha \tilde A]^{-1} \tilde a \tilde c^T.
\deq
And premultiplying (\ref{eq:InvBlock2}) by $[\tilde D - \alpha \tilde A]^{-1}$, we obtain
\beq
\label{eq:InvBlock2a}
\alpha [\tilde D - \alpha \tilde A]^{-1} \tilde a = \frac{1}{c_{n,n}} \tilde c.
\deq
Combining both equations (\ref{eq:InvBlock1a}) and (\ref{eq:InvBlock2a}) gives
$$
[\tilde D - \alpha \tilde A]^{-1} = \tilde C - \frac{1}{c_{n,n}} \tilde c \tilde c^T,
$$
and hence $
\varphi^{(1,n)} =  [\tilde D - \alpha \tilde A]^{-1} \tilde b = \tilde C_{\cdot,1} - \frac{c_{1,n}}{c_{n,n}} \tilde c$.
Thus, we can write
$$
\varphi^{(1,n)}_v-\varphi^{(1,n)}_w = (c_{v,1}-c_{w,1}) + \frac{c_{1,n}}{c_{n,n}} (c_{w,n}-c_{v,n})
$$
The above expression is symmetric and can be rewritten for any source-target pair $(s,t)$. That is,
$$
\varphi^{(s,t)}_v-\varphi^{(s,t)}_w = (c_{v,s}-c_{w,s}) + \frac{c_{s,t}}{c_{t,t}} (c_{w,t}-c_{v,t}).
$$
Furthermore, since matrix $C$ is symmetric for symmetric graphs, we can rewrite the above
equation as
$$
\varphi^{(s,t)}_v-\varphi^{(s,t)}_w = (c_{s,v}-c_{s,w}) + \frac{c_{s,t}}{c_{t,t}} (c_{t,w}-c_{t,v}),
$$
which completes the proof.
\mqed

\medskip

The current $I^{(s,t)}_e$ through edge $e=(v,w)$ is equal to $\alpha(\varphi^{(s,t)}_v-\varphi^{(s,t)}_w)$.
Let \[x_e^{(s,t)}=|\varphi^{(s,t)}_v-\varphi^{(s,t)}_w|,\quad (v,w)\in E\]
be the difference of potentials, that determines the absolute value of the current on the edge.
The \acfbet of edge $e$ is defined by
\begin{equation}
\label{eq:acfbet_edge}
x_e^\alpha=\frac{1}{n(n-1)}\sum_{s,t\in V, s\ne t}x_e^{(s,t)}, \quad e\in E.
\end{equation}
Further, for each node $v\in V$ its \acfbet is defined as the sum of the \acfbet scores of its adjacent edges:
\begin{equation}
\label{eq:acfbet}\mbox{\acfbet}(v)=\sum_{(v,w)\in E}x_{(v,w)}^\alpha,\quad v\in V.\end{equation}
With this definition, the node is central if a relatively large amount of current flows from this node to the network. This is in accordance to the original CF-betweenness of \cite{Newmana,Brandes2005a}, except we introduced the
additional sink ground node $n+1$. This mitigates the computational complexity because the original CF-betweenness require the inversion of the ill-conditioned matrix $[\tilde D - \tilde A]$, while for computing \acfbet we 
need to invert the matrix $[D - \alpha A]$, which is a well posed problem, and has many possible efficient 
solutions, for example, power iteration and Monte Carlo methods. In fact, as we shall show below, we need 
to obtain just a few rows of the inverse matrix $[D - \alpha A]^{-1}$.
In the rest of the paper we will discuss the computations and the properties of the \acfbet.

\section{Computation of \acfbet}
\label{sec:computation}

Due to the presence of the auxiliary node $n+1$, the value of $x_e^{(s,t)}$ on the right-hand side
of (\ref{eq:acfbet_edge}) can be computed efficiently with high precision for any source-destination pair.
However, the summation over all $n(n-1)$ pairs is a problem of prohibitive computational complexity
even for graphs of a modest size. The solution is to perform the computations for sufficiently many
source-destination pairs. This presents two problems: how to sample the source-destination pairs and
how many such pairs we need to achieve a good precision.

Ideally, we would like to choose the most representative source-destination pairs. In particular,
we can expect large values of $x_e^{(s,t)}$ if the sum of all potentials $\sum_{v\in V}\varphi^{(s,t)}_v$ is maximal.
Let us take again $s=1$, $t=n$. Then we obtain
\begin{align}
\label{eq:PageRank}
\sum_{v\in V}\varphi^{(1,n)}_v&={\bf 1}^T[\tilde D - \alpha \tilde A]^{-1}\tilde{b}
={\bf 1}^T[I - \alpha \tilde P]^{-1}\tilde{D}^{-1}\tilde{b},
\end{align}
where ${\bf 1}$ is a column vector of ones, and $\tilde P$ is the transition probability matrix
for a simple random walk on $G$ with absorption in $n$. Compare this to the well-known expression for
PageRank vector $\pi=(\pi_1,\ldots,\pi_n)$ with uniform teleportation and damping factor $\alpha$:
\[\pi=\frac{1-\alpha}{n}{\bf 1}^T[I-\alpha P]^{-1}.\]
Note that the vector ${\bf 1}^T[I - \alpha \tilde P]^{-1}$ in (\ref{eq:PageRank}) is very similar to PageRank,
except it nullifies the contribution of node $n$. We denote this vector by $\tilde{\pi}$ and recall that $\tilde{b}=(1,0,\ldots,0)^T$ to obtain
\[\sum_{v\in V}\varphi^{(1,n)}_v=\tilde{\pi}_1d_1^{-1}.\]
It is well-known and is also confirmed by our experiments that the PageRank of a node in an undirected graph
is strongly correlated to the degree of the node. Thus, with any choice of the source, the sum of the potentials is of similar magnitude, except for the cases when the contribution of the destination node is defining for the PageRank mass of the source. However, the destination node will mainly affect the PageRank of its close neighbours. Thus,
we propose to choose the source-destination pair uniformly at random, so that there is no preference on
the source, and the probability of choosing neighbour nodes is small. This results in the next algorithm
for computing the \acfbet.

\medskip

\noindent {\bf Algorithm~1}.
\begin{enumerate}
\item Select a set of pairs of nodes $(s_i,t_i), i=1,...,N$, uniformly at random;
\item For each $s_i$ or $t_i$, $i=1,...,N$ compute the rows $c_{s_i,\cdot}$, $c_{t_i,\cdot}$.
(this can be done either by power iteration or by Monte Carlo algorithm);
\item For each edge $e=(v,w)$ and each pair $(s_i,t_i)$, use (\ref{eq:potdifmodif}) to compute
$$
x^{(s_i,t_i)}_{e} = |\varphi_v-\varphi_w|.
$$
\item Average over source-destination pairs
$$
\bar{x}^{\alpha}_{e} = \frac{1}{N} \sum_{i=1}^N x^{(s_i,t_i)}_{e}.
$$
\end{enumerate}

Since we chose the pairs $(s_i,t_i)$ uniformly at random then for every edge $e$, $\bar{x}^{\alpha}_{e}$ is just
a sample average where all values are between zero and one. Then using the standard approach for the analysis
of the series of independent random variables we have the following result.

\begin{thm}\label{thm:complexity}
Algorithm~1 approximates the alpha current flow betweenness
in $\mbox{O}(m\log(n)\eps^{-2}\log(\eps)/\log(\alpha))$ time and $\mbox{O}(m)$ space
to within an absolute error of $\eps$ with arbitrarily high fixed probability.
\end{thm}

\noindent {\bf Proof:} In addition to the proof of Theorem~3 in \cite{Brandes2005a} we just need to note
that we can compute Personalized PageRank with precision $\eps$ in $\mbox{O}(\log(\eps)/\log(\alpha))$
power iterations.
\mqed

\medskip

%\noindent {\bf Algorithm~2}.
%\begin{enumerate}
%\item Run Algorithm-1;
%\item Choose top-k list of edges whose low confidence bound is larger than the computed average of alpha-current-flow-betweenness
%\item Run Algorithm-1 on the selected top-k list
%\end{enumerate}

\section{Truncated \acfbet}

In the experiments we noticed that the values $x_e^{(s,t)}$ have a high variance, which results in poor precision
when evaluating $x_e^{\alpha}$. A closer analysis revealed that the edges adjacent to the source $s$ receive large
values of $x_e^{(s,t)}$. This is especially apparent when $e=(v,s)$, where $v$ has degree~1, so $(v,s)$ is its only edge, and $s$ has a large degree. This can be explained using the random walk interpretation.
 Consider a PageRank-type random walk on $G$. At each node, with probability $\alpha$, the random walk traverses a randomly chosen edge of this node, and with probability $1-\alpha$ it jumps to the sink $n+1$. Denote by $T_B$ the number of steps of the random walk needed to hit set $B$.  Then it follows from Proposition~10 of \cite[Chapter 3]{Aldous-Fill} that $\varphi_v^{(s,t)}/\varphi_s^{(s,t)}=P_v(T_{\{s\}}<T_{\{t,n+1\}})$, where $P_v(\cdot)$ is a conditional probability
given that the random walk starts at $v$. Hence, if $s$ is the only neighbor of $v$ then $\varphi_v^{(s,t)}/\varphi_s^{(s,t)}=\alpha$, the probability of no absorption
 before reaching $s$. Thus, $|\varphi_s^{(s,t)}-\varphi_v^{(s,t)}|=(1-\alpha)\varphi_s^{(s,t)}$, which can be large if e.g. $\alpha=0.8$ because  $\varphi_s^{(s,t)}$ is the largest potential in the network. Furthermore, the original CF-betweenness corresponds to $\alpha=1$, implying that the current in $(v,s)$ is zero.

\begin{figure}[htb]
\centering
\includegraphics[width=.7\textwidth]{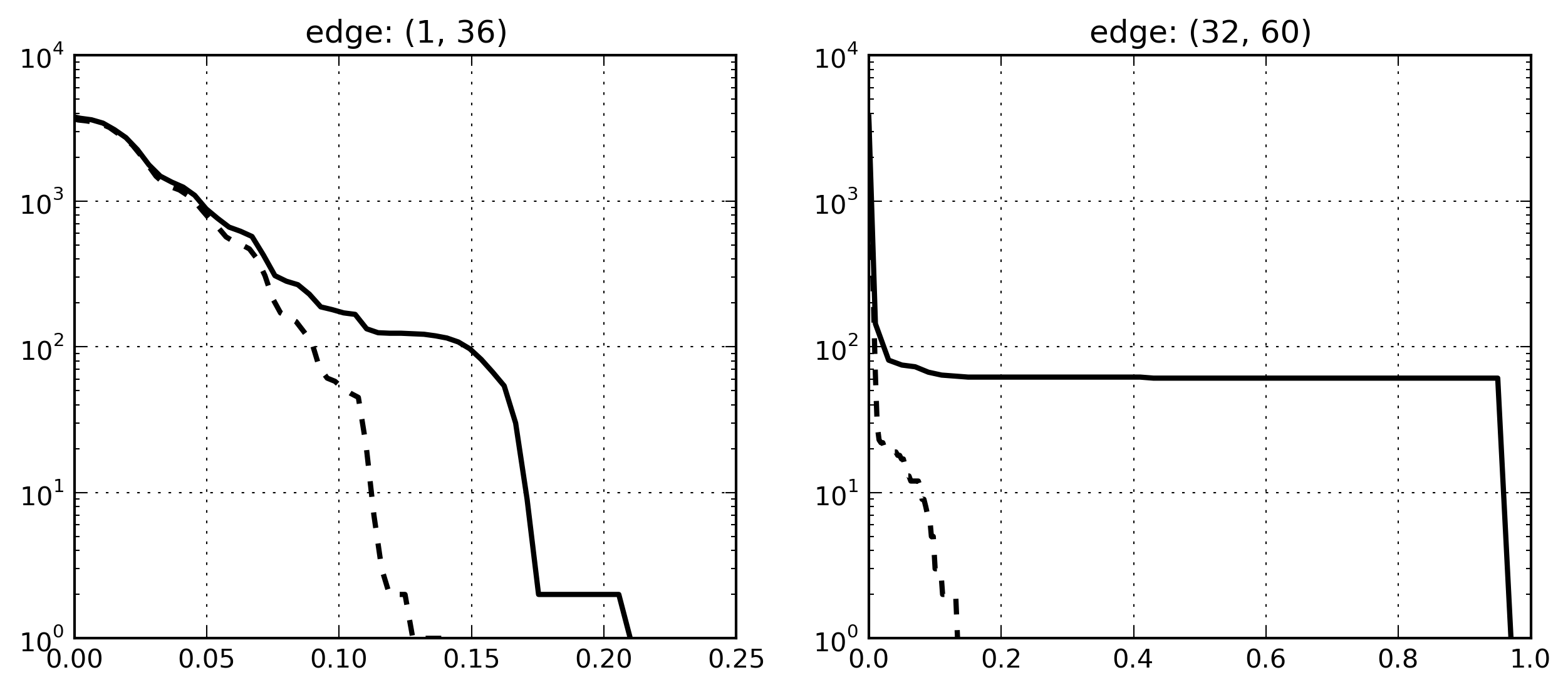}
{\caption{The number of pairs $s,t$ with $x_{(v,w)}^{(s,t)}>x$ over all pairs $(s,t)$ (solid line) and only pairs with $v,w\ne s$. (dashed line)}
\label{fig:truncated}}
\end{figure}

 This motivates for the truncated version of \acfbet where for each edge $(v,w)$ we only take into account the scores $x_{(v,w)}^{(s,t)}$
if $v,w\ne s$. In Figure~\ref{fig:truncated} we present log-linear plots of the empirical complementary distribution function of
$x^{(s,t)}_{(v,w)}$ over all pairs $(s,t)$ (solid line), and its truncated version (dashed line). The plots are given for two edges in the Dolphin social network described in Section~\ref{sec:datasets} below. Nodes 1 and 36 are central in the network, so the high \acfbet of (1,36) is expected. Node 60 has degree~1, so  edge (32,60) gains an unwanted high betweenness in the non-truncated version.
	
Since the truncated \acfbet gives lower scores to the edges connected to nodes of degree~1, one can expect that it has a higher correlation with CF-betweenness,  especially for not very large $\alpha$. This is confirmed below in Figure~\ref{fig:corr-truncated}. Moreover,  the truncated version removes outliers, and does not have large spread in values, thus standard statistical procedures, based on the Central Limit Theorem can be applied. Also, because of the smaller variance, Algorithm~1 achieves a desired precision with a smaller sample of source-destination pairs.

\section{Datasets}
\label{sec:datasets}

We consider the four graphs described below.

%	\item \textbf{Coauthorships in network science}

%	Nodes of coauthorships graph represent authors of articles in network science field while edges represent presence of joint publications \cite{Newman}.	
	
\textbf{Dolphin social network.} This small graph represents a social network of frequent associations between 62 dolphins in a community living off Doubtful Sound, New Zealand \cite{Lusseau2003}.		
	
 \textbf{Graph of VKontakte social network.}
	%\begin{figure}[htb]
%\centering
%\includegraphics[width=.5\textwidth]{vk_graph.png}
%\caption{Subraph of VKontakte social network}
%\label{fig:vkGraph}
%\end{figure}	
We have collected data from a popular Russian social network VKontakte. We were considering subgraph representing one of the connected components of people who stated that they were studying at Applied Mathematics - Control Processes Faculty at the St. Petersburg State University in different years. We ran the breadth-first search (BFS) algorithm starting at one specific node on the network and then anonymized the obtained users' data leaving only information about connections between people. Collected network consists of 2092 individuals out of total 8859 denoted the specified faculty in the Education field.
	
 \textbf{Watts-Strogatz model.}	As an artificial example, we used a random graph generated by the Watts-Strogatz model. We have chosen this model as it combines high clustering and short average path length, thus different centrality measures give very different results on this graph. For other random models considered (Erdos-Renyi and Barabasi-Albert) all measures are highly correlated and behave very similar to each other.
	
 \textbf{Enron graph.}	Enron email communication network is a well known test dataset. It covers all the email communication within a dataset of around half million emails between Enron's employees. The node are e-mail addresses, and the edges appears if an e-mail message was sent from one e-mail to another. Although this graph is small compared to, say, web or Twitter samples, it is already prohibitively large for computing the CF-betweenness in its original form.

\begin{table}[htbp]
\label{table:graphs}
\centering
\begin{footnotesize}
\begin{tabular}{l *6c }
\toprule
                                           		    &  $|V|$ &  $|E|$ &  $\langle deg(v) \rangle$ &  $diam(G)$ &  $C_\text{clustering}$ &  $\langle d(u,v) \rangle$ \\
\midrule
                             Dolphin social network 	&     62 &    159 &    5.13 &         8 &    0.259 &        3.357 \\
                        VKontakte AMCP social graph 	&   2092 &  14816 &   14.16 &        14 &    0.338 &        4.598 \\
              Watts-Strogatz 	&   1000 &   6000 &   12.00 &         6 &    0.422 &        3.713 \\
              ($n=1000, k=12, p=0.150$)&&&&&\\
              Enron & 36692& 183831 & 10.02 & 11& 0.4970 &$\approx$ 4.8\\
\bottomrule
\end{tabular}
\end{footnotesize}
\caption{Datasets characteristics}
\end{table}

\section{Numerical results for \acfbet}
\label{sec:correlations}

\begin{figure}[htb]
\centering
\includegraphics[width=.4\textwidth]{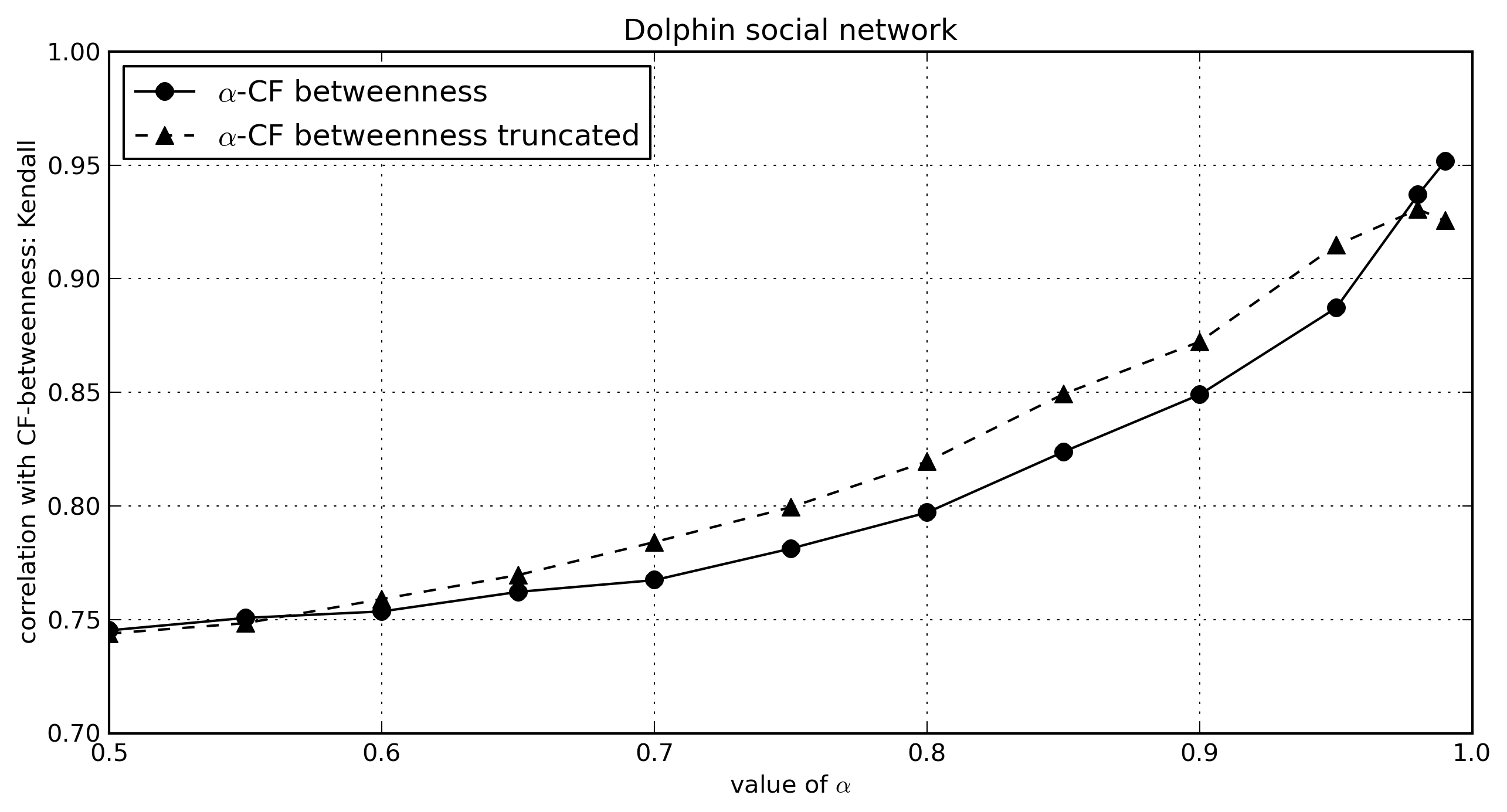}\hspace{.2cm}
\includegraphics[width=.4\textwidth]{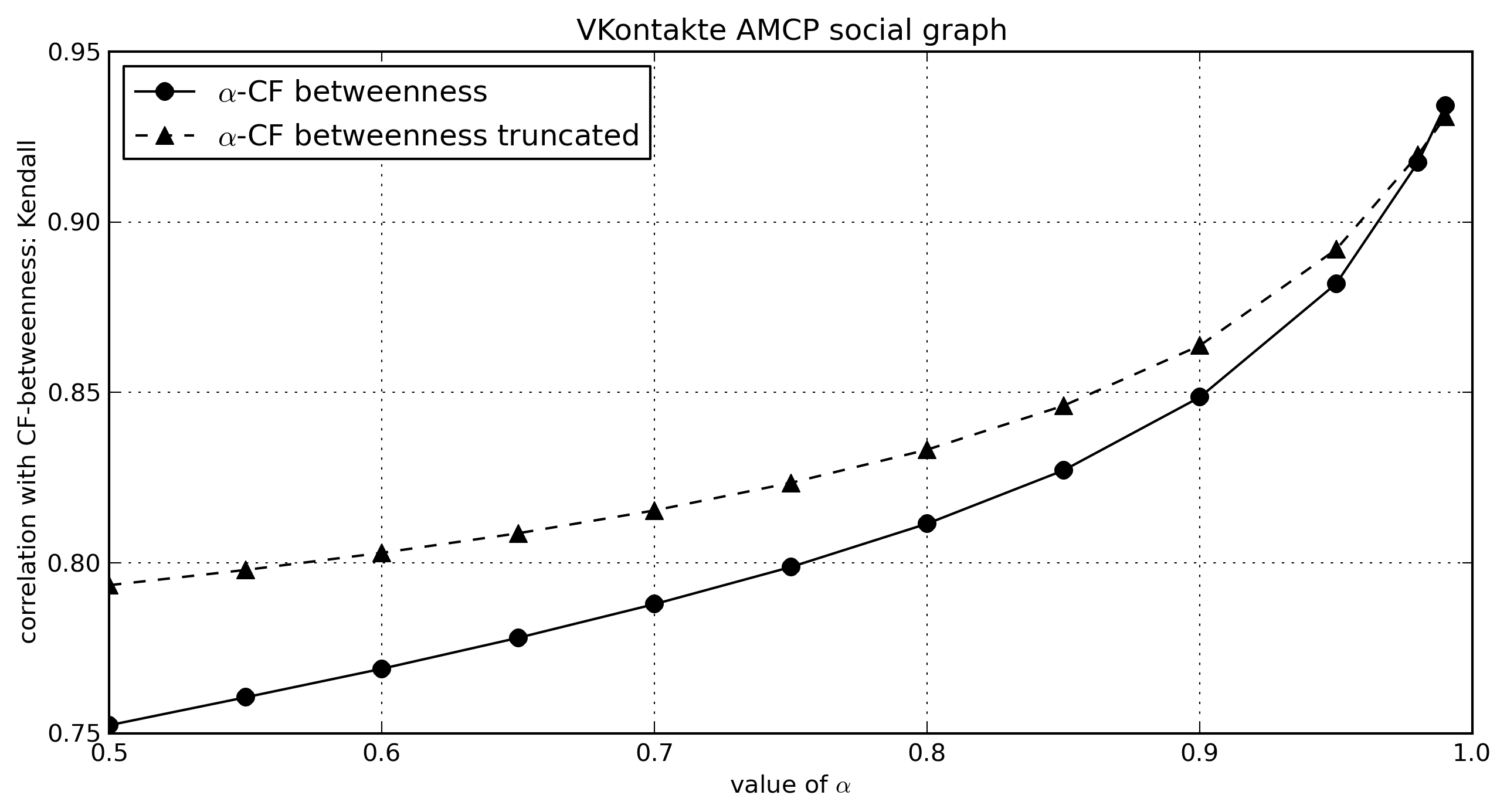}
\includegraphics[width=.4\textwidth]{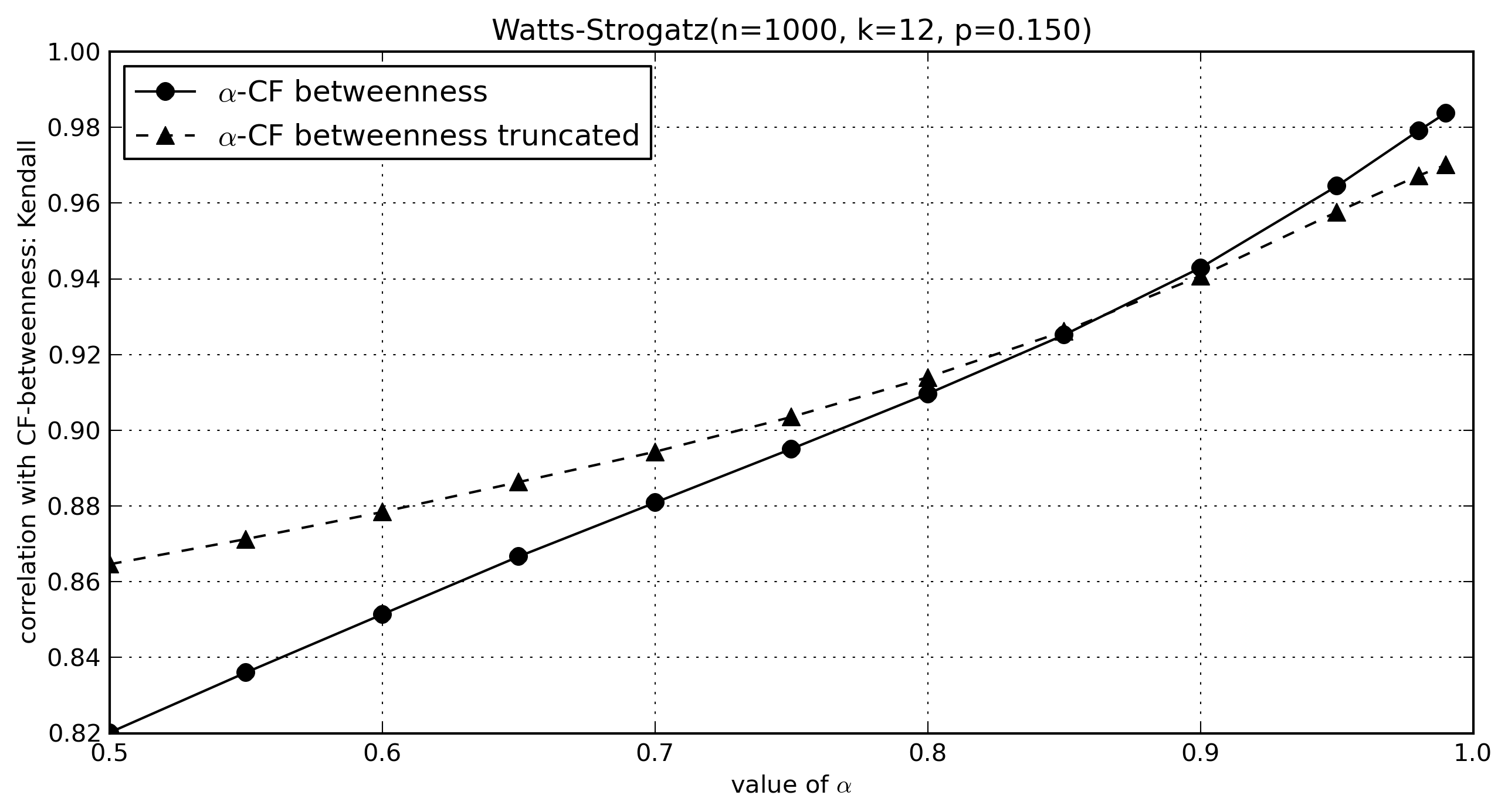}
{\caption{Correlations between \acfbet and truncated \acfbet with CF-betweenness as a function of $\alpha$.}
\label{fig:corr-truncated}}
\end{figure}

To begin with, we compare the two versions of \acfbet (truncated and without truncation) to the CF-betweenness scores defined as in \cite{Newmana,Brandes2005a}. Figure~\ref{fig:corr-truncated} presents the results for the three smaller graphs, in which the latter measure could be computed. As a correlation measure we use the Kendall tau rank correlation.
We  observe that the truncated version is better correlated with the CF-betweenness when $\alpha$ is not very close to one. As explained above, this is because the high probability of absorption results in a relatively high current in the edges connected to the source, which is not necessarily the case if absorption is only possible in the destination node.

Next, we demonstrate that that we can compute \acfbet in the Enron graph, where the computation of CF-centrality is infeasible. We have evaluated \acfbet, non-truncated and truncated, with $\alpha=0.98$. We have run Algorithm~1 using with $N=20\cdot 10^6$ source-destination pairs.
In the plot below we show the complementary distribution function in log-linear scale, of the score $x_e^{0.98}$ across the edges.
\begin{figure}[htb]
\centering
\includegraphics[width=.4\textwidth]{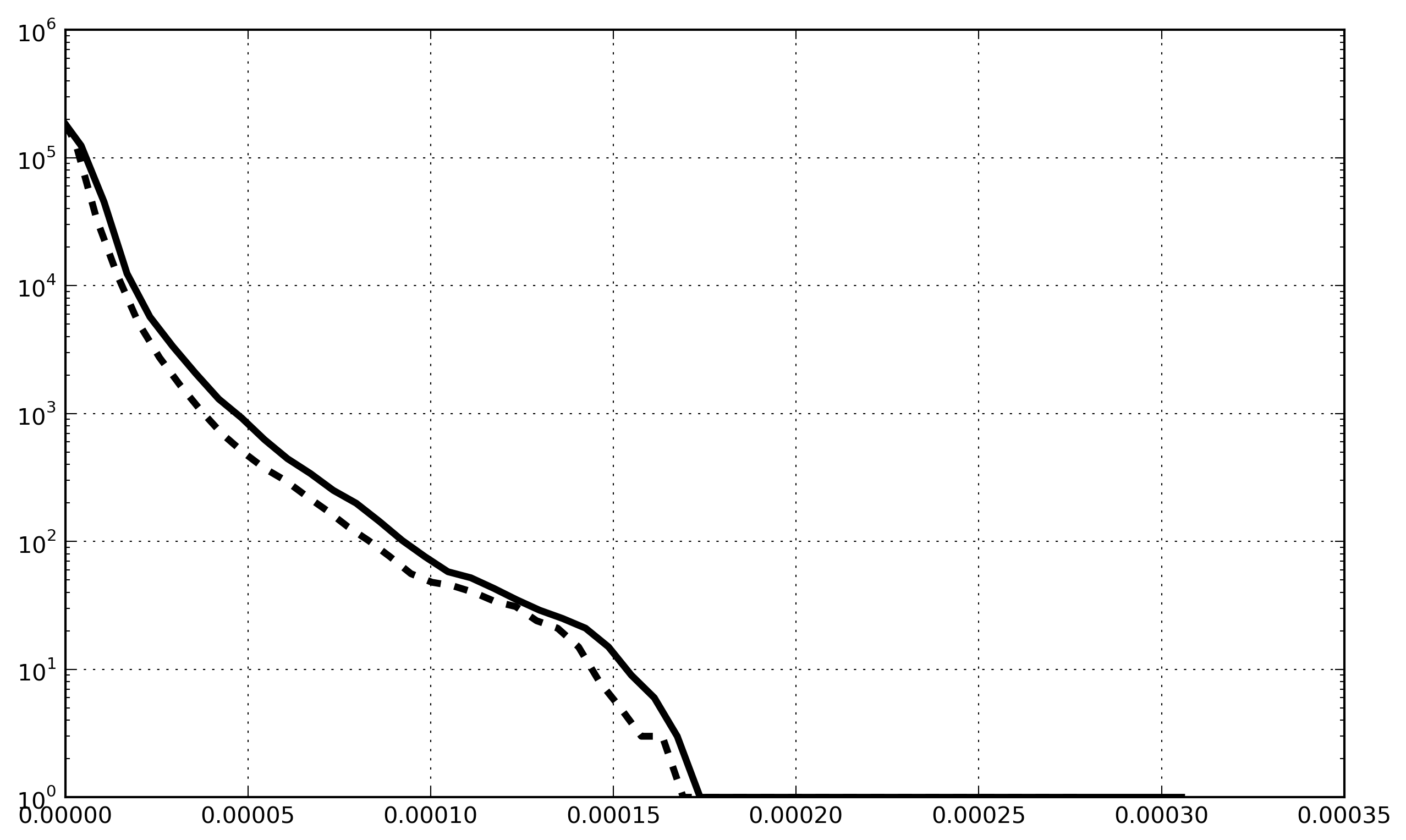}\hspace{.2cm}
\includegraphics[width=.4\textwidth]{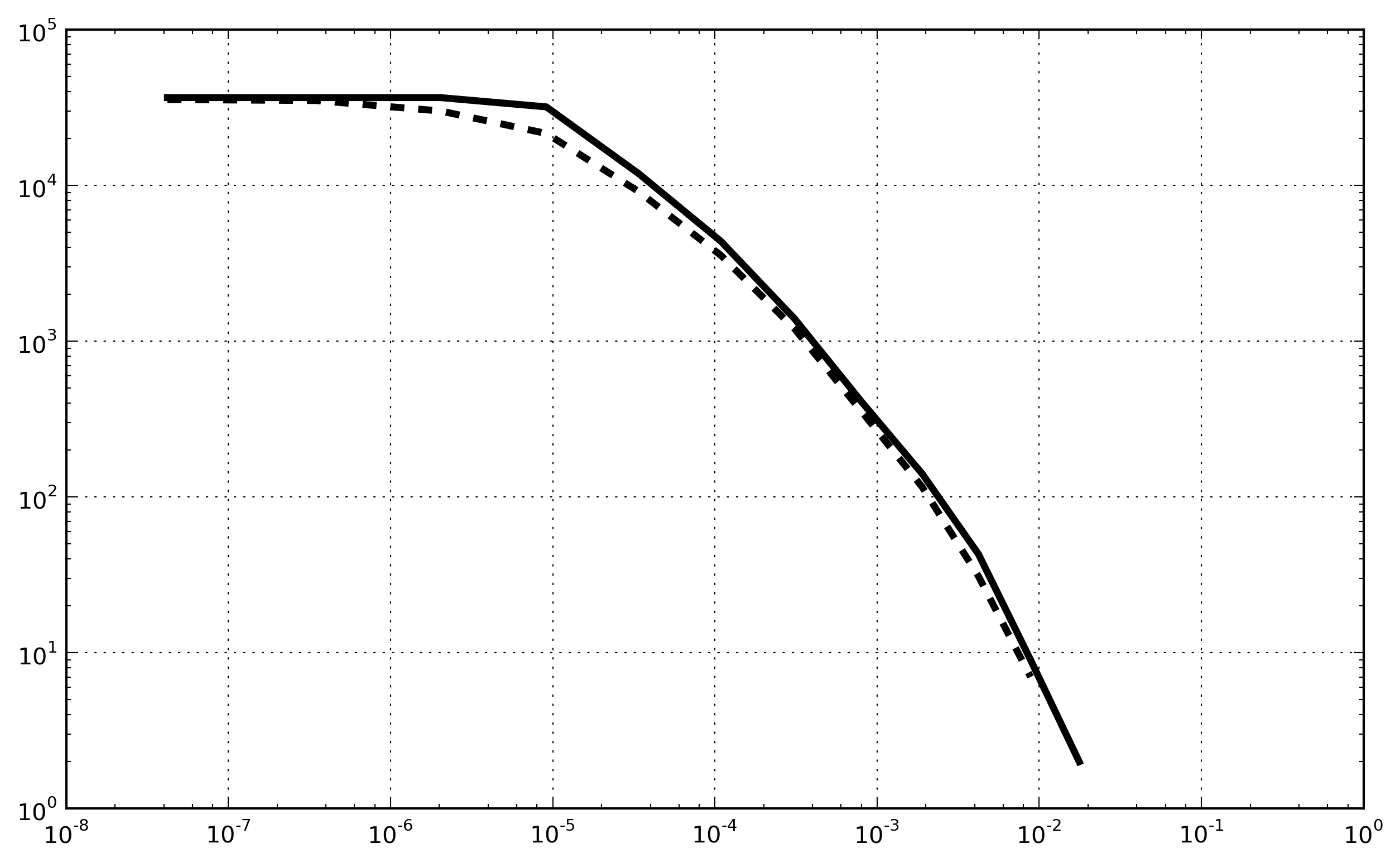}
{\caption{Distribution of \acfbet scores in the Enron graph, truncated (dashed line) and not truncated (solid line). Left: $x_e^{0.98}$ for edges $e\in E$. Right: \acfbet$(v)$ for $v\in V$. On the $x$-axis are the values of \acfbet, on the $y$-axis the number of edges/nodes with the score larger than $x$.}
\label{fig:enron}}
\end{figure}	

Note that distribution over edges (the left plot in Figure~\ref{fig:enron}) does not have a large spread of values, except one outlier edge that connects two most important hubs. Since the weights of the edges are comparable, it is to be expected that in this graph the nodes of large degrees are also the ones with highest betweenness. Indeed, the Kendall's tau correlation between  \acfbet and degree of the nodes turns out to be $0.808$, which is higher than in small examples below. The reason can be either the graph size or its structure. In future research we will investigate how the CF-betweenness score, e.g. its maximum value across the edges, scales with the graph size in graphs with power law degrees.

We further present correlations between our proposed measures and other measure of betweenness. These are computed on smaller graphs where we could obtain exact values of all presented measures, see Tables~\ref{tab:correlations-dolphins}--\ref{tab:correlations-ws}. For completeness, we also include one distance-base centrality measure - the Closeness Centrality: $$C_\text{C}(v)= \frac{n-1}{{\sum_{w\in V, w \neq v}d(v,w)}},$$
where $d(v,w)$ is the graph distance between $v$ and $w$. Betweenness (Between.) is computed as in (\ref{eq:sp-betweenness}), and PageRank(PR) is computed with $\alpha=0.85$.
{\small
\begin{table}[h]
\begin{tabular}{lcccccccc}
%\toprule
%\multicolumn{9}{l}{Dolphin social network}\\
\toprule
{} &  Degree &  PR &  Closeness &  Between. &  CF &  $\alpha$CF(0.8) &  $\alpha$CF-tr(0.8) &  $\alpha$CF(0.98) \\
\midrule
Degree             &   1.000 &     0.930 &      0.548 &        0.665 &         0.737 &            0.864 &               0.855 &             0.769 \\
PageRank           &   0.930 &     1.000 &      0.458 &        0.658 &         0.733 &            0.872 &               0.827 &             0.757 \\
Closeness          &   0.548 &     0.458 &      1.000 &        0.578 &         0.575 &            0.515 &               0.573 &             0.591 \\
Betweenness        &   0.665 &     0.658 &      0.578 &        1.000 &         0.829 &            0.749 &               0.759 &             0.828 \\
CF       &   0.737 &     0.733 &      0.575 &        0.829 &         1.000 &            0.798 &               0.820 &             0.939 \\
$\alpha$CF(0.8)    &   0.864 &     0.872 &      0.515 &        0.749 &         0.798 &            1.000 &               0.925 &             0.838 \\
$\alpha$CF-tr(0.8) &   0.855 &     0.827 &      0.573 &        0.759 &         0.820 &            0.925 &               1.000 &             0.876 \\
$\alpha$CF(0.98)   &   0.769 &     0.757 &      0.591 &        0.828 &         0.939 &            0.838 &               0.876 &             1.000 \\
\bottomrule
\end{tabular}
{\caption{Kendall tau for centrality measures in Dolphin social network.}
\label{tab:correlations-dolphins}}
\end{table}
\begin{table}[h]
\begin{tabular}{lcccccccc}
%\multicolumn{9}{l}{VKontakte AMCP social graph}\\
\toprule
{} &  Degree &  PR &  Closeness &  Between. &  CF &  $\alpha$CF(0.8) &  $\alpha$CF-tr(0.8) &  $\alpha$CF(0.98) \\
\midrule
Degree             &   1.000 &     0.655 &      0.679 &        0.521 &         0.545 &            0.659 &               0.668 &             0.599 \\
PageRank           &   0.655 &     1.000 &      0.375 &        0.662 &         0.717 &            0.833 &               0.811 &             0.766 \\
Closeness          &   0.679 &     0.375 &      1.000 &        0.382 &         0.356 &            0.424 &               0.445 &             0.395 \\
Betweenness        &   0.521 &     0.662 &      0.382 &        1.000 &         0.761 &            0.760 &               0.749 &             0.778 \\
Current Flow       &   0.545 &     0.717 &      0.356 &        0.761 &         1.000 &            0.812 &               0.833 &             0.917 \\
$\alpha$CF(0.8)    &   0.659 &     0.833 &      0.424 &        0.760 &         0.812 &            1.000 &               0.938 &             0.878 \\
$\alpha$CF-tr(0.8) &   0.668 &     0.811 &      0.445 &        0.749 &         0.833 &            0.938 &               1.000 &             0.903 \\
$\alpha$CF(0.98)   &   0.599 &     0.766 &      0.395 &        0.778 &         0.917 &            0.878 &               0.903 &             1.000 \\
\bottomrule
\end{tabular}
{\caption{Kendall tau for centrality measures in the social graph VKontakte AMCP.}
\label{tab:correlations-vk}}
\end{table}
\begin{table}[h]
\begin{tabular}{lcccccccc}
%\multicolumn{9}{l}{Watts-Strogatz(n=1000, k=12, p=0.150)}\\
\toprule
{} &  Degree &  PR &  Closeness &  Between. &  CF &  $\alpha$CF(0.8) &  $\alpha$CF-tr(0.8) &  $\alpha$CF(0.98) \\
\midrule
Degree             &   1.000 &     0.891 &      0.462 &        0.526 &         0.610 &            0.643 &               0.581 &             0.612 \\
PageRank           &   0.891 &     1.000 &      0.415 &        0.485 &         0.565 &            0.610 &               0.546 &             0.567 \\
Closeness          &   0.462 &     0.415 &      1.000 &        0.655 &         0.613 &            0.647 &               0.666 &             0.628 \\
Betweenness        &   0.526 &     0.485 &      0.655 &        1.000 &         0.853 &            0.819 &               0.852 &             0.857 \\
Current Flow       &   0.610 &     0.565 &      0.613 &        0.853 &         1.000 &            0.910 &               0.914 &             0.979 \\
$\alpha$CF(0.8)    &   0.643 &     0.610 &      0.647 &        0.819 &         0.910 &            1.000 &               0.935 &             0.923 \\
$\alpha$CF-tr(0.8) &   0.581 &     0.546 &      0.666 &        0.852 &         0.914 &            0.935 &               1.000 &             0.930 \\
$\alpha$CF(0.98)   &   0.612 &     0.567 &      0.628 &        0.857 &         0.979 &            0.923 &               0.930 &             1.000 \\
\bottomrule
\end{tabular}
{\caption{Kendall tau for centrality measures in the Watts-Strogatz graph (n=1000, k=12, p=0.150).}
\label{tab:correlations-ws}}
\end{table}
}

Note that \acfbet is strongly correlated with CF-betweenness. The Closeness Centrality does not agree well with the CF-betweenness, even the PageRank and the degrees have a higher correlations with the CF-betweenness in real graphs. Recent paper~\cite{Boldi2013axioms} suggests more measures based on distance, and efficient computation method for such measures is presented in \cite{Boldi2013hyperball}. In future it will be interesting to compare these new measures to \acfbet.

\section{Centrality measures and network vulnerability}

\begin{figure}[htb]
\centering
\includegraphics[width=.4\textwidth]{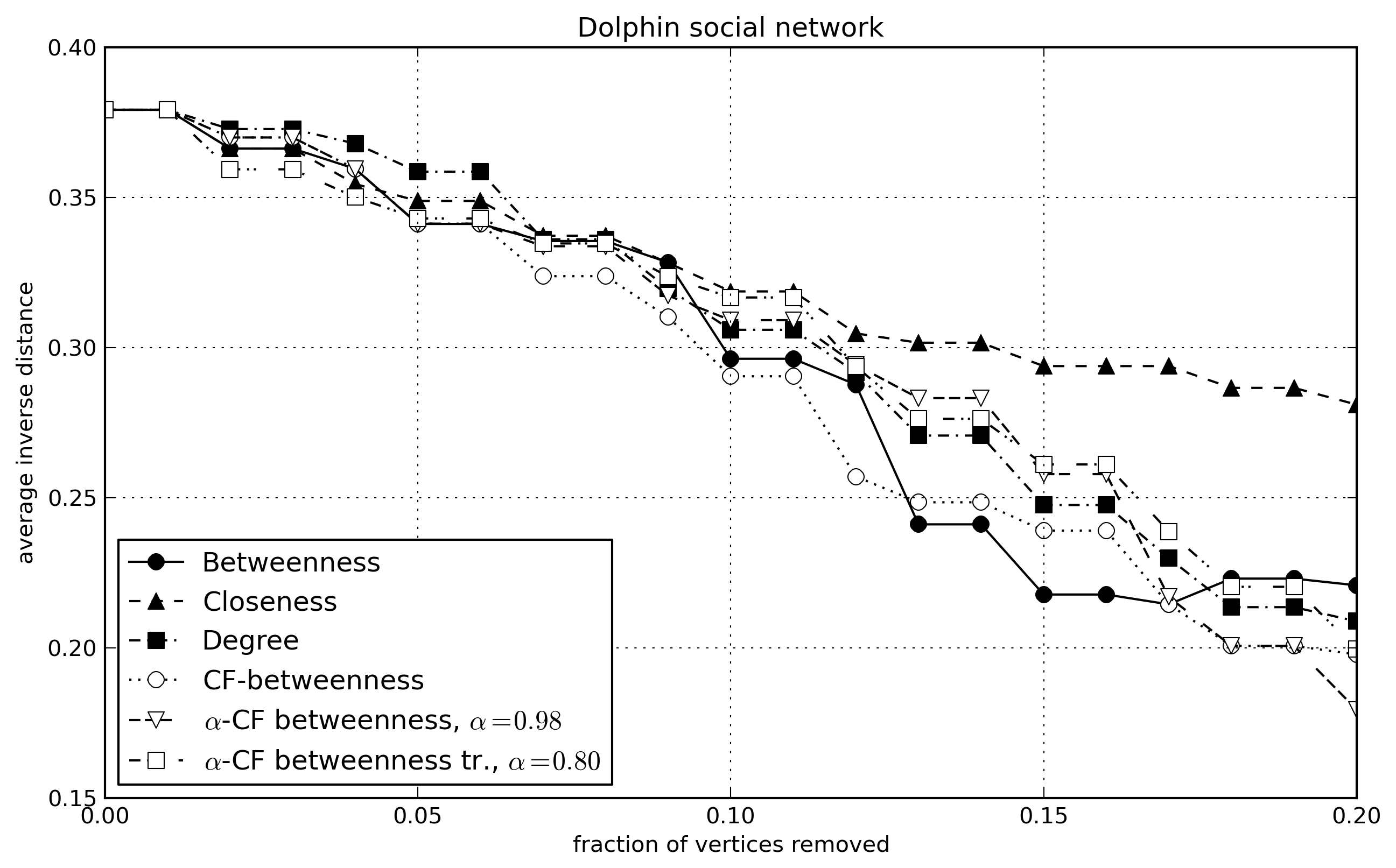}\hspace{.1cm}
\includegraphics[width=.4\textwidth]{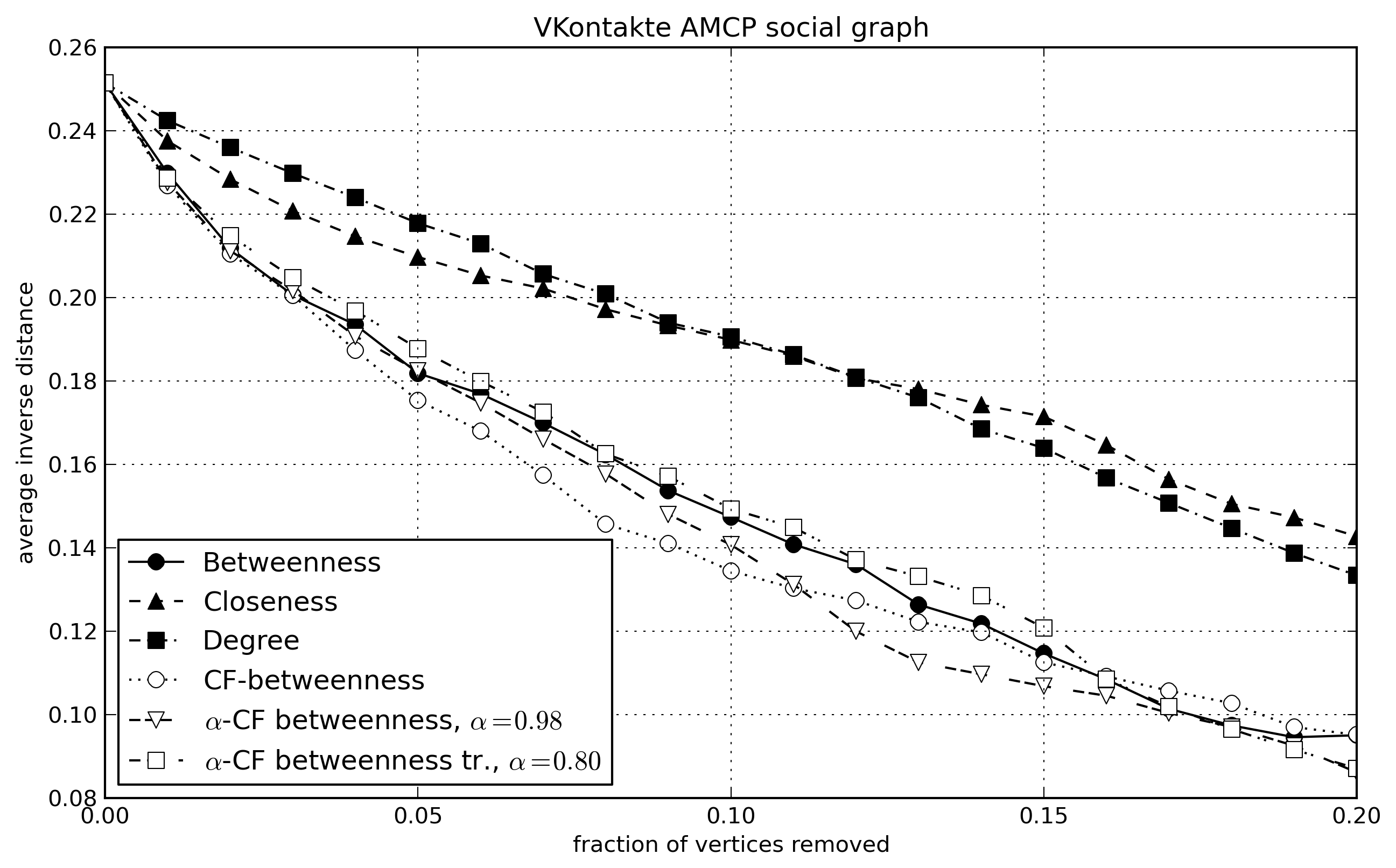}
\includegraphics[width=.4\textwidth]{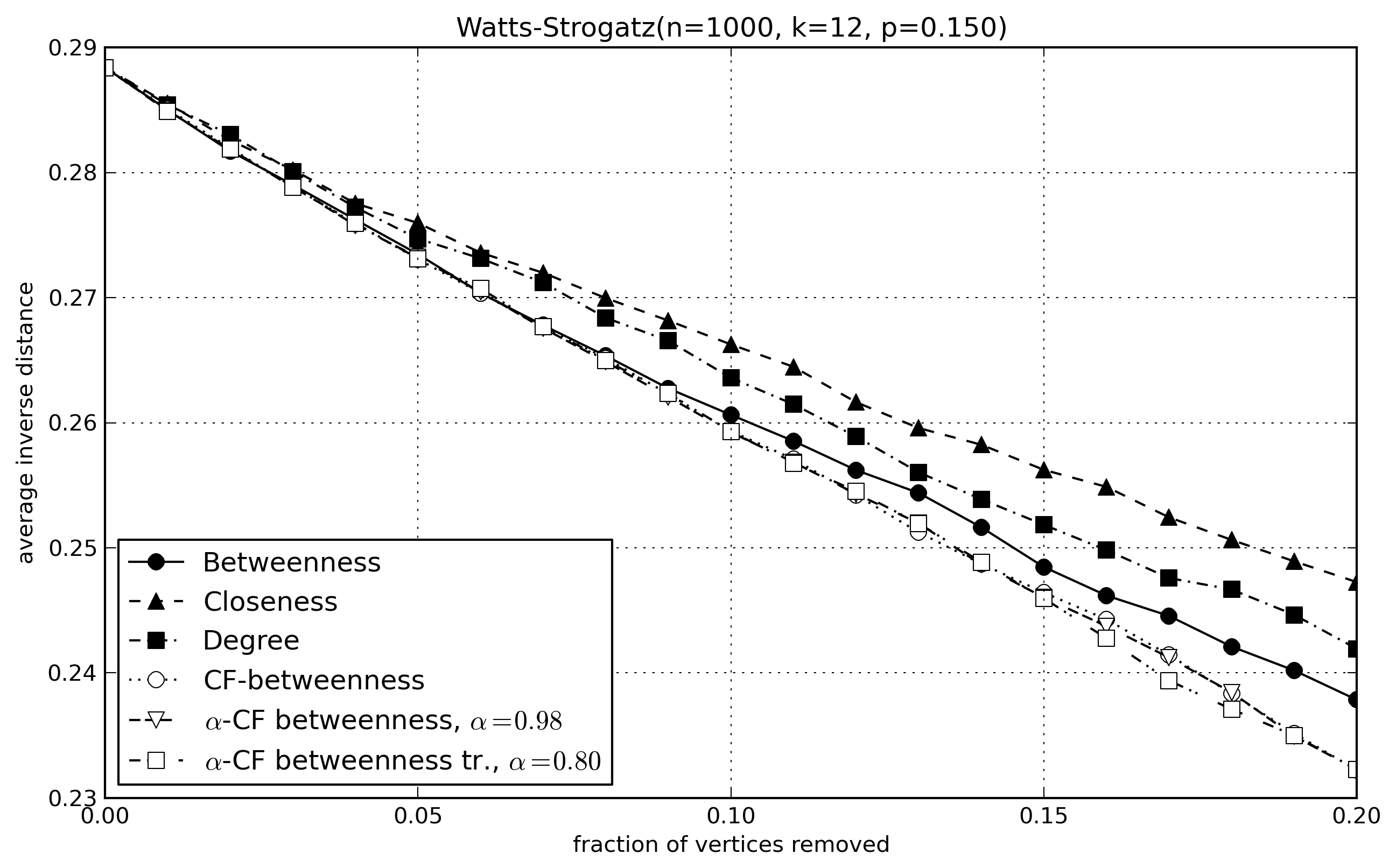}
{\caption{Inverse average distance as a function of the fraction of removed top-nodes according to different betweenness centrality measures.}
\label{fig:aid}}
\end{figure}

We now consider how well the CF-betweenness and \acfbet can indicate the nodes responsible for maintaining the network connectivity. We follow the methodology in \cite{Holme2002}. As measures of connectivity we choose the average inverse distance
\[<d^{-1}>=\frac{1}{n(n-1)}\sum_{u,v\in V, u\ne v}\frac{1}{d(u,v)}\]
and the size of the largest connected component. In the experiment, we remove the top nodes one by one, according to different betweenness measures,  and observe how the connectivity of the network changes.
In Figure~\ref{fig:aid} the results are presented for the inversed average distance.

The results for the social graph VKontakte are especially interesting, because this network turns out to be less vulnerable to the removal of nodes with large degree than nodes with large betweenness and its modifications (CF-betweenness, \acfbet, and truncated \acfbet). On the small Dolphin social network there is no much difference in vulnerability with respect to different centrality measures. Finally, on the artificial Watts-Strogatz graph the CF-betweenness and our proposed two versions of \acfbet find the nodes that are most essential for the network connectivity.

\begin{figure}[htb]
\centering
\includegraphics[width=.4\textwidth]{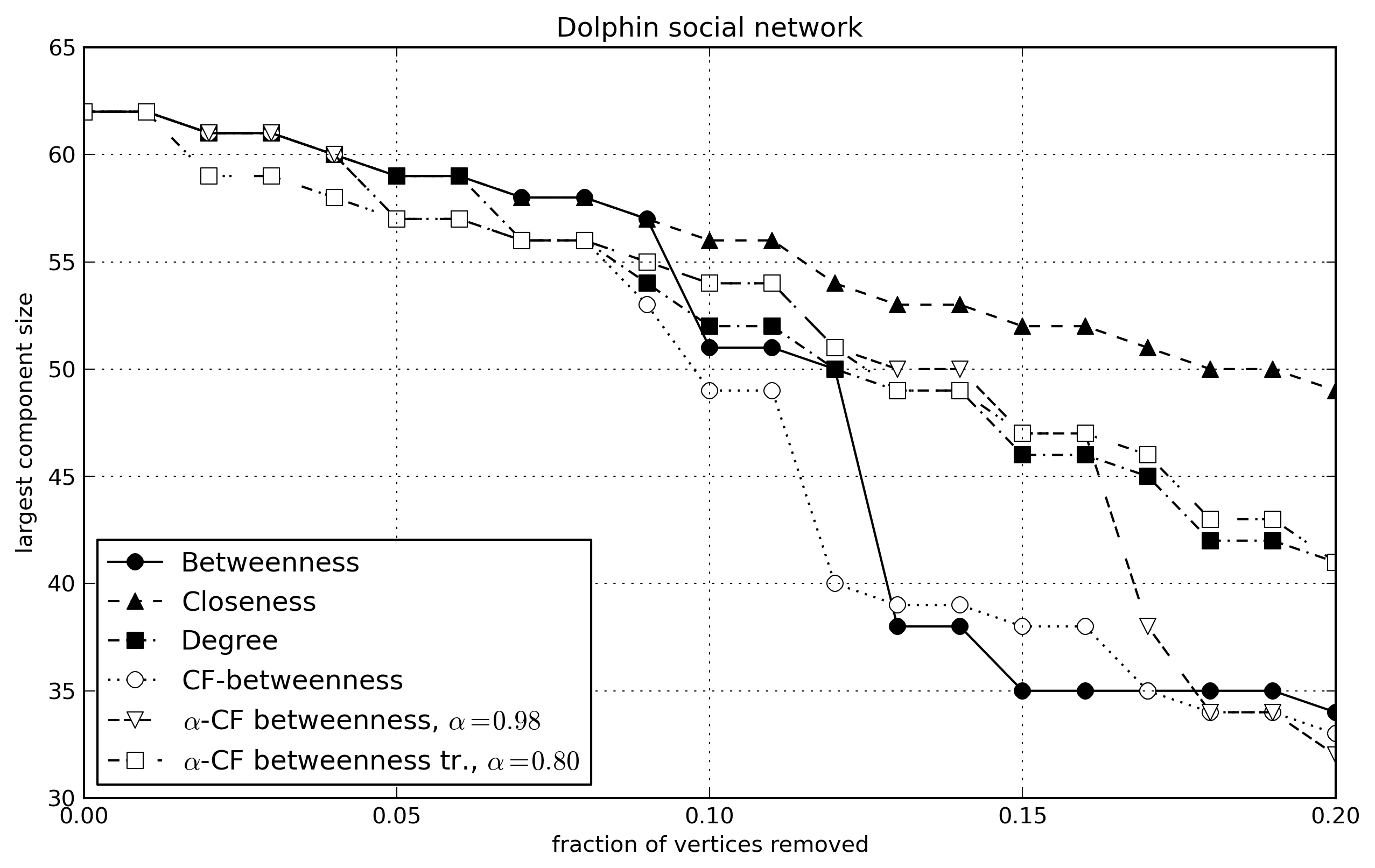}\hspace{.1cm}
\includegraphics[width=.4\textwidth]{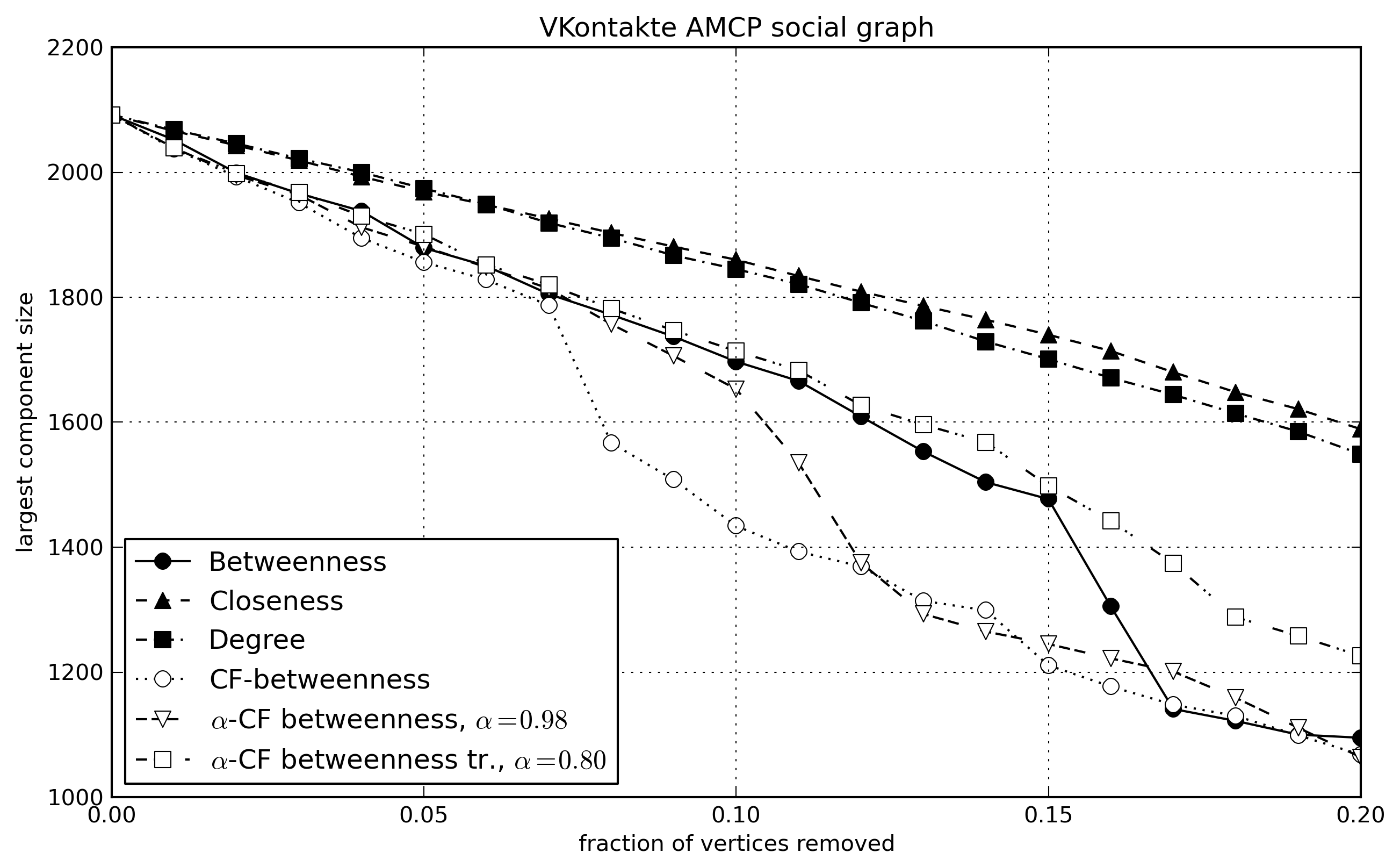}
{\caption{The size of the largest connected component as a function of the fraction of removed top-nodes according to different betweenness centrality measures.}
\label{fig:lcs}}
\end{figure}

Another connectivity measure of the network is the size of its larges connected component. In Figure~\ref{fig:lcs} we plot the size of the largest connected components against the fraction of removed top-nodes. We do not present the plot for the Watts-Strogatz graph because it remains entirely connected, so the size of its largest connected component equals to the number of remaining nodes irrespectively of which nodes are removed first. For the two real graphs, the CF-betweenness is most efficient in reducing the size of the giant component. On the Dolphin graph, \acfbet performs closely to CF-betweenness, except the interval when 13-18\% of nodes are removed. On the graph VKontakte, \acfbet and its truncated version perfom comparably to the CF-betweenness. Again, on this graph, degree and Closeness centrality fail to reveal the nodes responsible for the network connectivity.
The \acfbet with $\alpha=0.98$ appears to be a better measure for betweenness of a node than the truncated \acfbet with $\alpha=0.8$. The latter however also gives gives good results, and  can be computed easier on large graphs due to the faster convergence of the power iteration algorithm. 

We conclude that both \acfbet and truncated \acfbet provide an adequate measure for the role of a node in network's connectivity. Furthermore, their computational costs are lower than for known measures of betweenness, and the computations can be done in parallel easily. Thus, \acfbet can be applied in large graphs, for which computations of other measures of betweenness are merely infeasible.

\bibliographystyle{plain}
\bibliography{library}

\begin{thebibliography}{10}

\bibitem{Aldous-Fill}
D.~Aldous and J.~Fill.
\newblock Reversible {M}arkov chains and random walks on graphs.
\newblock 1999.

\bibitem{Boldi2013axioms}
P.~Boldi and S.~Vigna.
\newblock Axioms for centrality.
\newblock {\em arXiv:1308.2140}.

\bibitem{Boldi2013hyperball}
P.~Boldi and S.~Vigna.
\newblock In-core computation of geometric centralities with hyperball: A
  hundred billion nodes and beyond.
\newblock {\em arXiv:1308.2144}.

\bibitem{Brandes2001}
U.~Brandes.
\newblock {A faster algorithm for betweenness centrality}.
\newblock {\em Journal of Mathematical Sociology}, 25(1994):163--177, 2001.

\bibitem{Brandes2005a}
U.~Brandes and D.~Fleischer.
\newblock {Centrality measures based on current flow}.
\newblock In {\em Proceedings of the 22nd annual conference on Theoretical
  Aspects of Computer Science}, pages 533--544, 2005.

\bibitem{Brin1998107}
S.~Brin and L.~Page.
\newblock {The anatomy of a large-scale hypertextual Web search engine}.
\newblock {\em Computer Networks and \{ISDN\} Systems}, 30(1–7):107--117,
  1998.

\bibitem{DoyleSnell}
P.G. Doyle and J.L. Snell.
\newblock {\em Random walks and electric networks}.
\newblock Mathematical Association of America, 1984.

\bibitem{Freeman1977}
L.~C. Freeman.
\newblock {A set of measures of centrality based on betweenness}.
\newblock {\em Sociometry}, 1977.

\bibitem{Holme2002}
P.~Holme, B.J. Kim, C.N. Yoon, and S.K. Han.
\newblock {Attack vulnerability of complex networks.}
\newblock {\em Physical review. E, Statistical, nonlinear, and soft matter
  physics}, 65(5 Pt 2):056109, May 2002.

\bibitem{Lusseau2003}
D.~Lusseau, K.~Schneider, O.J. Boisseau, P.~Haase, E.~Slooten, and S.M. Dawson.
\newblock {The bottlenose dolphin community of Doubtful Sound features a large
  proportion of long-lasting associations}.
\newblock {\em Behavioral Ecology and Sociobiology}, 54(4):396--405, September
  2003.

\bibitem{Newmana}
M.E.J. Newman.
\newblock {A measure of betweenness centrality based on random walks}.
\newblock {\em Social networks}, pages 1--15, 2005.

\end{thebibliography}

\end{document}